\documentclass[a4paper,12pt]{article}
\usepackage{latexsym}
\usepackage{amsmath,amssymb,bm,ascmac,color,calc}
\usepackage[mathscr]{eucal}
\usepackage{graphicx}
\usepackage{cite}
\usepackage{lineno}
\usepackage{setspace}
\usepackage{etoolbox}
\usepackage{url}
\usepackage{longtable} 
\usepackage{multirow,booktabs}
\usepackage[affil-it]{authblk}
\allowdisplaybreaks 

\begin{document}

\begin{titlepage}

\title{{\large\bf Search for axion-like particles with electron and positron beams at KEK Linac}\vspace{1.2em}}

\author[1,3,4]{\small Akimasa~Ishikawa\thanks{akimasa.ishikawa@kek.jp}}
\author[2,3]{\small Yasuhito~Sakaki\thanks{sakakiy@post.kek.jp}}
\author[1,3]{\small Yosuke~Takubo\thanks{yosuke.takubo@kek.jp}\vspace{1.2em}}

\affil[1]{\small Institute of Particle and Nuclear Studies, High Energy Accelerator Research Organization (KEK), Ibaraki 305-0801, Japan.}
\affil[2]{\small Radiation Science Center, High Energy Accelerator Research Organization (KEK), Ibaraki 305-0801, Japan.}
\affil[3]{\small The Graduate University for Advanced Studies (SOKENDAI), Hayama 240-0193, Japan.}
\affil[4]{\small International Center for Elementary Particle Physics, the University of Tokyo, Tokyo 113-0033, Japan.}

\date{}

\maketitle

\vspace{1cm}

\begin{abstract}
We propose a fixed-target experiment to search for 
Axion-like particles (ALPs) coupling to photons, which utilizes electron and positron beams available at KEK Linac. The sensitivity to ALP is evaluated for two shielding setups in conjunction with other beam dump experiments, fixed-target experiments, and $e^+e^-$ collider experiments. 
This study shows that the two setups can explore the gap between the search region excluded by the beam dump experiments and the $e^+e^-$ collider experiments.
\end{abstract}

\thispagestyle{empty}

\end{titlepage}

\setlength{\parskip}{10pt}%

\section{Introduction}

In recent years, various non-collider experiments to search for new particles in physics beyond the Standard Model have been proposed~\cite{Battaglieri:2017aum, Beacham:2019nyx}. 
One of them is the fixed target experiment, whose sensitivity to new particles is complementary to that of the collider experiment. 
In the fixed target experiment, new particles can be generated via a variety of physics processes like meson decays and photon bremsstrahlung with large cross-section, realizing high sensitivity to new particles in the MeV-GeV range.

Each fixed target experiment is characterized by the type of beam particles, the bunch charge and the decay process of the new particles produced.  
Electrons, positrons, protons, photons, and muons are most commonly used as beam particles, which induce different physical processes in the target. 
The bunch charge and beam extraction scheme are relevant to the amount of the background occurring simultaneously in one bunch ({\it pileup}), and the final signal statistics.  
A capability of searches for invisible decays strongly depends on such beam conditions. 
For example, 
NA64($\mu$)~\cite{Gninenko:2016kpg,Banerjee:2016tad,Chen:2017awl,NA64:2018iqr}, 
MMAPS~\cite{Alexander:2017rfd}, 
LDMX~\cite{Berlin:2018bsc,Akesson:2018vlm} and 
positron beam experiments at Jefferson Lab~\cite{Marsicano:2018glj,Batell:2021ooj} use appropriate beam currents or bunch charges to reduce pileup and enable the search for invisible decays.

In this paper, we propose a fixed target experiment using 7\,GeV electron and 4\,GeV positron beams available in KEK Linac~\cite{Akemoto:2013lxa}, investigating its sensitivity to new physics, where we consider Axion-like particles (ALPs) as a benchmark.
The KEK Linac produces high bunch charge beams mainly for SuperKEKB, Photon Factory (PF) and PF Advanced Ring (PF-AR)~\cite{Akai:2018mbz}. 
We studied experimental setups to search for a visible decay of ALP with high statistics by using the beam, where a large background due to the high bunch charge should be dealt with by appropriate shielding.

We focus on parameter regions with shorter lifetimes than those explored in previous beam dump experiments. 
In this region, short-lived new particles tend to decay in the shield, then this region could not be explored in previous beam dump experiments with long shields.  
A short experimental setup is therefore needed to explore the region of interest, and SeaQuest experiment~\cite{Berlin:2018pwi} using proton beams is one good example.  
We evaluate sensitivity to ALP in the experiment using electron and positron beams, taking into account the background contamination.

This paper is organized as follows.
In Sec.~2, the ALP model is introduced. 
In Sec.~3, we describe the experimental setup, in particular the geometry of the shielding system, the beam conditions, and the detector concept.
In Sec.~4, details of background and signal estimation are explained.
In Sec.~5, the sensitivity to ALP for the experimental setups is shown. 
Section 6 is devoted to the summary.

\section{Model}
We consider ALPs described by the following effective Lagrangian:
\begin{align}
\delta\mathcal{L} =
	-\frac{1}{4}g_{a\gamma\gamma} a F_{\mu\nu}\tilde{F}^{\mu\nu}
	+\frac{1}{2}(\partial_{\mu}a)^2
	-\frac{1}{2}m_a^2a^2,
	\label{eq:L_ALP}
\end{align}
where $a$ is the ALP, $F^{\mu\nu}$ is a strength of the photon field, and $\tilde{F}^{\mu\nu}=\epsilon_{\mu\nu\lambda \rho}F^{\lambda\rho}/2$.
In our evaluation, it is assumed that a coupling $g_{a\gamma\gamma}$ and a ALP mass $m_a$ are independent parameters.

\begin{figure}[!t]
\begin{center}
\includegraphics[width=10.0cm, bb=0 0 820 200]{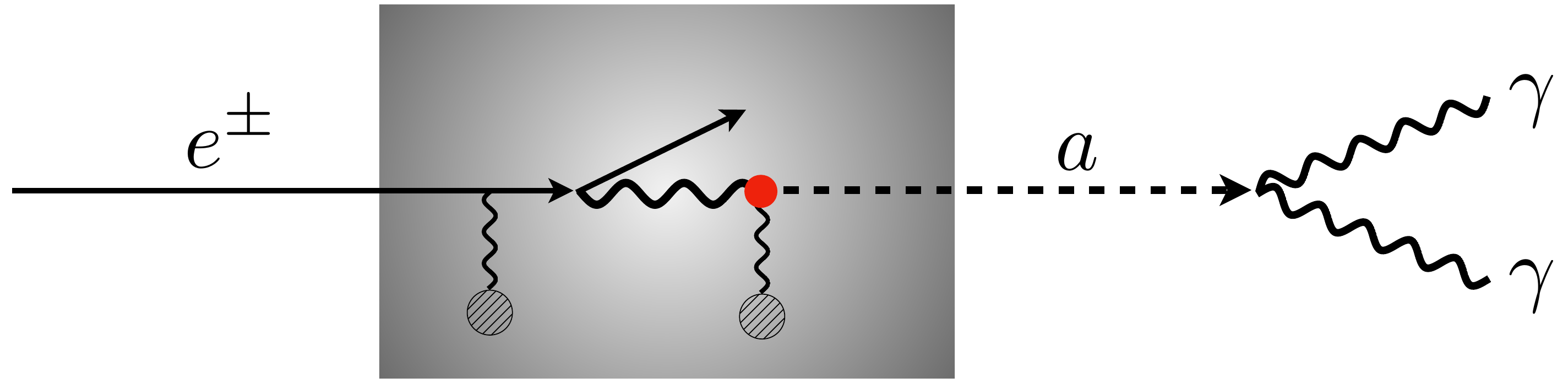}
\caption{{\footnotesize Schematic diagram of the production of ALP ($a$) by the Primakoff process and their decay behind a shield.}}
\label{fig:exp}
\end{center}
\end{figure}

In assuming this model, the ALP is mainly produced by photon-nucleus interactions, the Primakoff process, in the target as shown in Fig.\,\ref{fig:exp}. After passing through the shield, the ALP decays into two photons that reach the detector and are observed as a signal.

For the ALP production, we use the following angular differential cross section~\cite{Tsai:1986tx, Bjorken:1988as, Dobrich:2015jyk, Dusaev:2020gxi}, based on the improved Weizsacker-Williams approximation~\cite{vonWeizsacker:1934nji,Williams:1935dka,Kim:1973he}, 
\begin{align}
\frac{d\sigma_{\gamma a}}{d\theta_a}
	\simeq \frac{\alpha g_{a\gamma\gamma}^2 k^4\theta_a^3}{4t^2} G_2(t),
\end{align}
where $G_2(t) \simeq Z^2
	\left(a'^2t/(1+a'^2t)\right)^2/
	\left(1+t/d\right)^2$, 
$Z$ is the atomic number of target, 
$a'=112 Z^{-1/3}/m_e$, $d=0.164\text{ GeV}^2$, and $m_e$ is the electron mass. 
The momentum transfer $t$ is approximated as $t\simeq k^2\theta_a^2+m_a^4/(4k^2)$ to avoid cancellation of significant digits in numerical calculations~\cite{Dusaev:2020gxi}. 
For the form factor $G_2$, we use the combined simple atomic and nuclear form factor, which takes into account the screening effect. The decay width of the ALP is given by
\begin{align}
\Gamma_a=\frac{g_{a\gamma\gamma}^2 m_a^3}{64\pi}.
\end{align}
These formulae is adopted to calculate the signal yield.

\section{Experimental setups}
This section describes the geometries of the shielding system, beam conditions, and detector concept. We propose two experimental setups ({\it Setup-1} and {\it-2}) with different shielding geometries. In both setups, the maximum size is 5-10\,m in the beam axis direction and 1-2\,m in the vertical axis direction, which can be installed in the east beamline of the third beam switchyard at KEK Linac facility~\cite{Iwase:2020}.

\subsection{Geometries of the shielding system}
In the range of ALP mass from MeV to GeV, most of the small coupling regions of ALP $(g_{a\gamma\gamma} \lesssim 10^{-5,-6})$ have already been explored, mainly by a beam dump experiment with a very high number of incident particles (SLAC-E137~\cite{Bjorken:1988as}) and by the SN1987A supernova cooling~\cite{Jaeckel:2017tud,Dolan:2017osp}. 
For that reason, we focus on the larger coupling region, in other words, the {\it shorter lifetime region}. 
To increase sensitivity to the region, the shield needs to be shortened to avoid ALP decaying within the shield. 
In the following, we consider two experimental setups with as short shield as possible.

\subsubsection{Setup-1: Almost zero background}
\begin{figure}[!t]
\begin{center}
\includegraphics[width=14.0cm, bb=0 0 822 312]{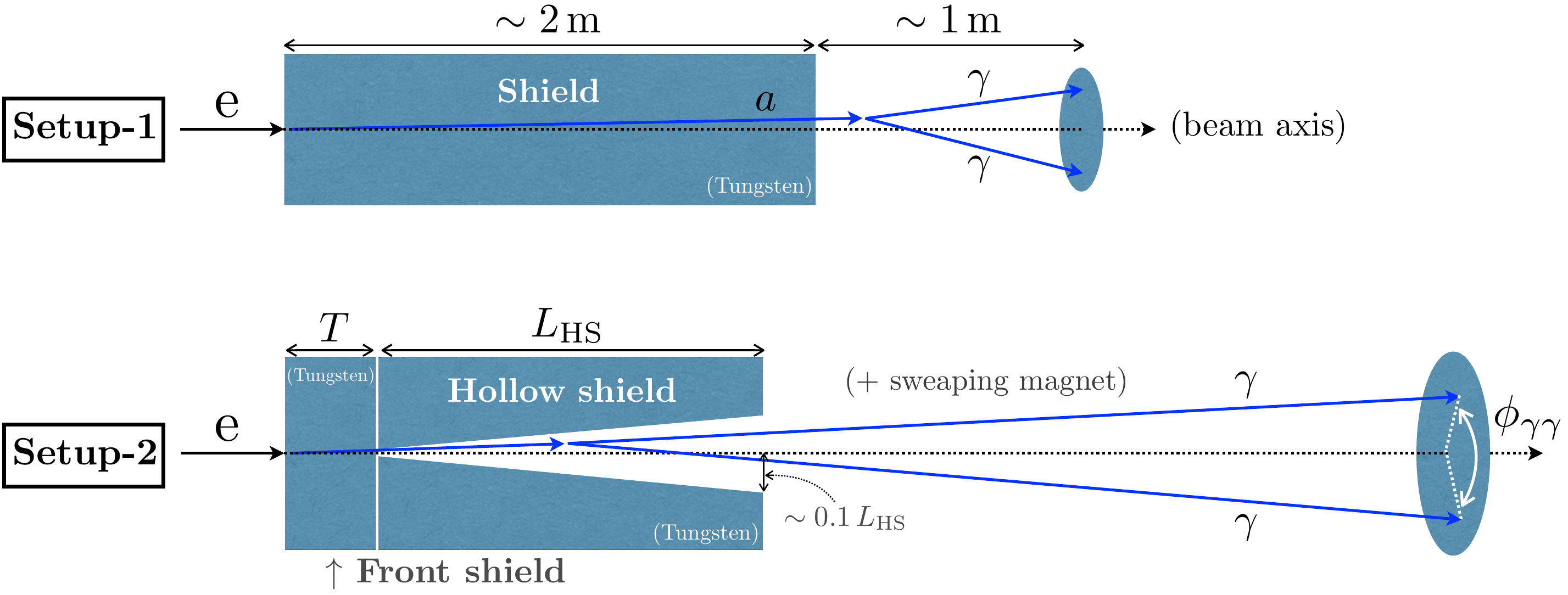}
\caption{{\footnotesize 
Schematic diagram of two shield systems Setup-1 and Setup-2.  
For Setup-1, a 2\,m tungsten shield is placed and 1\,m decay volume is behind it.  
For Setup-2, it has two shields, {\it Front shield} and {\it Hollow shield}.  The thickness of Front shield is $T=20\text{\,-\,}25$\,cm.  The hollow shield is hollowed out in the shape of a truncated cone. 
}}
\label{fig:setup}
\end{center}
\end{figure}

We first consider a simple setup, referring to as {\it Setup-1}. Figure \ref{fig:setup} (top) shows a schematic diagram of the setup. A 2\,m tungsten shield is placed and 1\,m decay volume is behind it. A similar setup has been seen in previous experiments~\cite{Konaka:1986cb,Bjorken:1988as,Davier:1989wz}.  

This setup can reduce the beam-related background to a negligible level for high-energy signals that we are interested in. The main sources of background are high-energy muons and neutral hadrons.
A material with high density efficiently works as their absorber and tungsten is sufficient for that purpose.

In Setup-1, we use positron beams with the lower energy of 4\,GeV. 
In this zero background setup, both the ALP decay length and rate of background muons penetrating the shield are proportional to the beam energy, therefore, the sensitivity to the shorter lifetime region is almost independent of the beam energy.
We adopted the lower energy beam, allowing us to remove the muons with a shorter shield and accordingly reduce the total cost of the experiment.

\subsubsection{Setup-2: With backgrounds}

The second setup ({\it Setup-2}) has two shields, 
{\it Front shield} and {\it Hollow shield}, as shown in Fig.\,\ref{fig:setup} (bottom).
The feature of Setup-2 is that the Front shield is thinner than that of Setup-1. 
Although the background increases with the thinner shield, the sensitivity to the shorter lifetime region improves with respect to Setup-1 due to the exponential increase of the signal.

In Setup-2, we use electron beams with the higher energy of 7\,GeV. 
The number of ALP passing through the shield exponentially becomes larger by using a higher energy beam without a significant increase in the background.
This is because the decay length of ALP is proportional to the beam energy, while the energy dependence of the attenuation length of neutral hadrons, the main background, is small.

We set the thickness of Front shield to $T=20\text{-}25$\,cm (about 60-70 radiation length).
The shield reduces the energy deposition or radiation damage to the detector, mainly due to the electromagnetic showers. 
The leakage of the neutral hadrons and the low-energy components of the electromagnetic showers cannot be prevented by this Front shield completely. 
A Hollow shield is placed just behind the Front shield, which is hollowed out in the shape of a truncated cone. 
The longitudinal length of Hollow shield is set to $L_{\rm HS} = 100$\,cm.
The upstream side facing the Front shield has a smaller radius of the truncated cone of $0.01\,T$ while the radius of the downstream side is set to $0.1\,L_{\rm HS}$. 
This shape is efficient to reduces the leakage particles from Front shield since they are produced at a wider angle than the signal.
ALP passing through the Front shield is emitted approximately on the beam axis. The ALP or decayed photons pass through the center (hollow part) of the Hollow shield, and the signal photons can reach the detector.

For brevity, this study assumes that charged particles can be removed by a sweeping magnet placed behind the Hollow shield. 
About 2\,T$\cdot$m is enough to bend charged particles with approximately beam energy by 0.1 radians without hitting the detector. 
The detector is located 5\,m downstream from the upstream surface of Front shield with 50\,cm in radius.

The above shield parameters are tentative and their optimization depends on the configuration of the detector, the available beam parameters, and the decay particles in the signal, which we will leave as future works of great interest.

\subsection{Beam conditions}
The beam conditions assumed in this study are summarised in Table~\ref{table:beam}. 
To evaluate the sensitivity in both experimental setups, the number of electrons and positrons incident on the target is set to $6.5\times 10^{18}$, which corresponds to the beam current supplied to the SuperKEKB for one month.
The sensitivity in Setup-2 depends on the bunch charge of the beam, and we use two cases with 0.2\,nC and 4\,nC as benchmarks. The former is the typical bunch charge supplied by KEK Linac to PF and PF-AR, and the latter is that to be used in the final phase of SuperKEKB.

The typical beam current supplied by Linac for SuperKEKB is 400\,nA with the bunch charge of 0.2\,nC, 40 bunches per pulse, and a pulse repetition frequency of 50\,Hz.
If the bunches are evenly spaced in 4\,$\mu$s, a typical RF pulse width of KEK Linac, the bunch spacing is 100\,ns. 
For 4\,nC bunch, the bunch spacing becomes longer than that value due to the smaller number of bunches per pulse.

\begin{table}[!t]
  \centering
  \caption{\footnotesize Beam conditions assumed in this study.}
  \vspace{7pt}
  \label{table:beam}
  \begin{tabular}{ccccc} 
  \toprule[1pt]
     & Beam
     & Energy
     & Total number of particles
     & Bunch charge (density) \\ 
  \toprule[0.5pt]
  Setup-1
     & $e^+$
     & 4\,GeV
     & $6.5\times 10^{18} $
     & -  \\ 
  \toprule[0.5pt]
  \multirow{2}{*}{Setup-2}
     & \multirow{2}{*}{$e^-$}
     & \multirow{2}{*}{7\,GeV}
     & \multirow{2}{*}{ $6.5\times 10^{18} $ }
     & 0.2\,nC ~$(1.25\times 10^{9}\,{\rm e})$ \\
     &
     &
     &
     & 4.0\,nC ~$(2.5\times 10^{10}\,{\rm e})$  \\
  \toprule[1pt]
  \end{tabular}
\end{table}

\subsection{Detector concept}

For detection of photons from an ALP decay, we assume multi-layers of sampling ElectroMagnetic (EM) calorimeter consisting of an active detector and tungsten absorber plate with about 3.5 mm thickness.

The tungsten plate creates electromagnetic shower which is detected by the active detector behind it. 
An example of the active detector is a scintillator with a Silicon Photo Multiplier (SiPM) readout, where SiPM mounted on a Printed Circuit Board (PCB) is housed in a round cavity in the center of a scintillator tile with size of 2\,cm\,$\times$\,2\,cm\,$\times$\,3\,mm. This detector concept is the same as Analogue Hadron sampling CALorimter (AHCAL) developed in CALICE collaboration \cite{Chau:2018ycb}. If the detector is assumed to be disk with 50\,cm radius, the number of readout channel is about 2,000 per layer.

The scintillator with SiTCP readout has a timing resolution better than 1~ns with respect to 100\,ns of the beam bunch space, therefore, hits originated from the same beam bunch can be identified. The timing information is useful to reject backgrounds with two photons originated in different bunches.

The analog-to-digital conversion can be performed in between two consecutive beam bunches by using a flash ADC. The energy resolution is expected to be better than 15\%/$\sqrt{E/\mathrm{GeV}}$, considering the number of electron-positron pairs generated in electromagnetic shower in the tungsten plate and using flash-ADC.

\section{Background and signal estimations}

\subsection{Setup-1}
In Setup-1, only the signal is considered since the background contamination is negligible. 
The method of evaluating the number of signals is summarised in detail in Ref.~\cite{Sakaki:2020mqb,Asai:2021ehn}.  The number of signals is given by
\begin{align}
N_{\rm signal} &= 
	\int_{m_a}^{E_{\rm beam}} dE_{\gamma}
	\int_{0}^{\pi} d\theta_a
	\int_{0}^{l_{\rm dec}} dz \notag\\
	&\times
	N_{\rm beam}\,
	n_{\rm atom}
	\frac{dl_\gamma}{dE_{\gamma}}
	\, \frac{d\sigma_{\gamma a}}{d\theta_a}
	\, \frac{dP_{\rm dec}}{dz}
	\, \Theta(r_{\rm det}-r_{\perp})
	   \Theta(E_a-E_{\gamma\gamma}^{\rm (cut)}),
	\label{eq:nsig_axion}
\end{align}
where 
$N_{\rm beam}$ is the number of beam particles incident on the target, 
$n_{\rm atom}$ is the number density of tungsten atoms in the target, 
$dl_\gamma/dE_{\gamma}$ is the track length of the photon in the target, 
$d\sigma_{\gamma a}/d\theta_a$ is the angular differential cross section of ALP, 
$dP_{\rm dec}/dz$ is the decay probability of ALP,
$z$ is the coordinate in the beam axis direction, starting at the rear of the shield and 
$l_{\rm dec}$ is the decay volume length. 
The first Heaviside step function represents a detector angular acceptance, where $r_{\rm det}$ and $r_{\perp}$ are the radius of the detector and the deviation of the signal photon from the beam axis.
The second Heaviside step function represents a detector energy acceptance, where $E_a$ and $E_{\gamma\gamma}^{\rm (cut)}$ are ALP energy and a cutoff of photon energy sum.  We assume $r_{\rm det}=15$\,cm and $E_{\gamma\gamma}^{\rm (cut)}=1$\,GeV in our analysis for Setup-1.

\subsection{Setup-2}
In Setup-2, the main background is due to $K^0$ and neutrons produced in photonuclear reactions from hard bremsstrahlung. 
In particular, $K^0_S$ passing through the Front shield and Hollow shield decays almost 100\% in front of the detector and makes a significant contribution to the background. 
We note that the decay length of $K^0_S$ with a kinetic energy of 6.3\,GeV (the maximum value when using a 7\,GeV beam) is 37\,cm, which is much smaller than the distance between the shields and the detector. 
Neutrons, $K^0_L$ and other hadrons can produce $\pi^0$ and $\eta$ through hadronic interactions in the Front shield, Hollow shield, and the absorption layer of electromagnetic calorimeters, which can also produce background photons.

To estimate these backgrounds, we performed Monte-Carlo simulations using {\tt PHITS}~\cite{Sato:2018}\footnote{
We also used {\tt Geant4}~\cite{Agostinelli:2002hh} and {\tt FLUKA}~\cite{Battistoni:2015epi}, but {\tt PHITS} showed no unphysical structure in the distribution of photoproduced hadrons, and reproduced the experimental data on photonuclear reaction relatively well.}.
We found that the Monte-Carlo code underestimates the amount of $K^0$ production, whose cross-section  was corrected in the code based on the experimental data~\cite{Boyarski:1969iy}\footnote{
We found that for both {\tt PHITS} and {\tt Geant4}, the Monte-Carlo prediction tends to underestimate the forward cross-section of $K$-meson as the energy of the initial photon increases, so we have corrected the code based on experimental data. The experimental values are for a charged $K$-meson rather than a neutral one, but their cross sections are similar at high energies~\cite{Egorov:2020ghz}.}.
For the neutron photoproduction, we confirmed that the Monte-Carlo reproduces well the experimental data~\cite{Anderson:1969jw} at incident photon energies around 4\,GeV, therefore, do not correct the Monte-Carlo code.

For the evaluation of the signal, we implement the Primakoff process as well as the decay process of the ALP described in Sec.\,2 in {\tt PHITS} and then perform Monte-Carlo simulations. 
The accuracy of the equations for the Primakoff process used is at $\mathcal{O}(\alpha)$ level. 
The flux of the initial photons in the electromagnetic shower is simulated by {\tt EGS5}~\cite{Hirayama:2005zm} embedded in {\tt PHITS} with an accuracy of a few percent or less~\cite{Nelson:2006te}. 
The signal is therefore expected to be predicted with higher accuracy compared to the background.

Our detector simulation and analysis are performed as follows:
\begin{itemize}
 \item Assuming a cell size of 2\,{\rm cm}\,$\times$~2\,{\rm cm}, energy of photons passing through the same cell in the same bunch is simply combined, which we call a {\it cluster}. The cluster position is assigned to its center.
 \item Smear the cluster energy with the assumed energy resolution $\sigma_E = 0.15 / \sqrt{ E_{\gamma}/{\rm GeV}}$.
 \item Discard the clusters with energy below 1\,GeV.
 \item Select the hardest cluster $(i=1)$, create a pair with it and an other softer cluster $(i\geq2)$, and calculate the energy sum $(E_{\rm tot}=E_1+E_i)$, invariant mass $(m_{\gamma\gamma})$ and azimuthal angle $(\phi_{\gamma\gamma})$ for them.
 For single cluster events, it is set to $E_{\rm tot}=E_1$, and $m_{\gamma\gamma}$ and $\phi_{\gamma\gamma}$ are not defined nor used for cut. 
\end{itemize}

Taking into account all photons coming from upstream, the hit occupancy in EM calorimeter with the segmentation of 2\,cm\,$\times$\,2\,cm is expected to be maximally 1\% with 1\,GeV threshold even at the innermost region with radius of 0-10~cm for 0.2\,nC and 4.0\,nC of beam bunch charge with $T=20\text{-}25\,$cm. For that reason, a signal hit can be well separated from background hits, making the event reconstruction with the above criteria feasible.

To calculate the mass, we need to define the momentum vector of each photon. 
Assuming that the photon is generated at the downstream surface of Front shield on the beam axis, we define the momentum vector pointing to the cluster position from that of the photon generation.
The azimuthal angle is defined as the angle between the position vectors of the two clusters on the detection plane, where the intersection of the beam axis and the detection plane is selected as the origin (see Fig.\,\ref{fig:setup}).

The selection cuts using these variables are summarized as
\begin{itemize}
 \item $E_{\rm tot}>6$\,GeV,
 \item $\phi_{\gamma\gamma}>2.5$,
 \item the mass window cut depending on ALP masses.\footnote{The width of the mass window is the two standard deviations in the Gaussian fit of the mass distribution.}
\end{itemize}
The significance is calculated as the root sum square of the significances for the multiple and the single cluster event category.

\begin{figure}[!t]
\begin{center}
\includegraphics[width=9.0cm, bb=0 0 424 320]{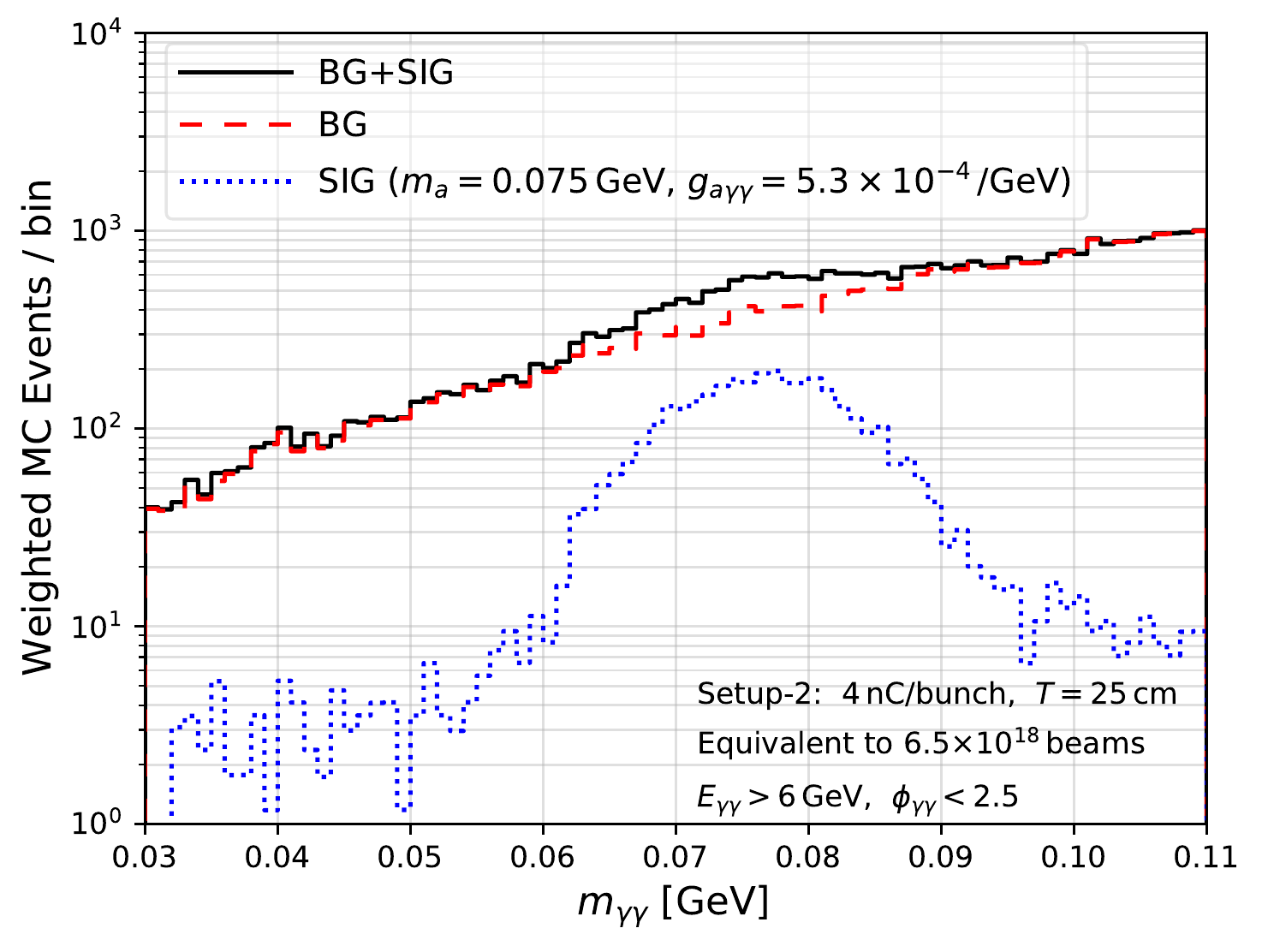}
\caption{{\footnotesize 
The $m_{\gamma\gamma}$ distribution of the signal and background after imposing $E_{\gamma\gamma}$ and $\phi_{\gamma\gamma}$ cuts.
The red dashed and blue dotted lines show the background and signal distributions respectively, and the black solid line shows the sum of the two. 
The bunch charge is assumed to be 4\,nC and the thickness of the Front shield is 25\,cm.
The parameter of the signal is $(m_a,g_{a\gamma\gamma})=(0.075\,{\rm GeV}, 5.3\times 10^{-4}{\rm /GeV})$.
}}
\label{fig:mass}
\end{center}
\end{figure}
Figure~\ref{fig:mass} is the $m_{\gamma\gamma}$ distribution of the background and the signal after imposing the $E_{\gamma\gamma}$ and $\phi_{\gamma\gamma}$ cuts.
The bunch charge is assumed to be 4\,nC and the thickness of the Front shield is set to 25\,cm.
The signal creates a clear bump for the signal with 
$(m_a,g_{a\gamma\gamma})=(0.075\,{\rm GeV}, 5.3\times 10^{-4}{\rm /GeV})$.

For the background study, it would be possible to employ the region of small-$\phi_{\gamma\gamma}$, and the sideband region of ALP mass as control regions. 
Data for several shield thicknesses $T$ would also be useful since the $T$-dependence of the attenuation lengths in the shield for the neutral hadrons and ALP are different. They would also work when the ALP mass is close to the $\pi^0$ mass.

\section{Results}

\begin{figure}[!t]
\begin{center}
\includegraphics[width=7.0cm, bb=0 0 300 286]{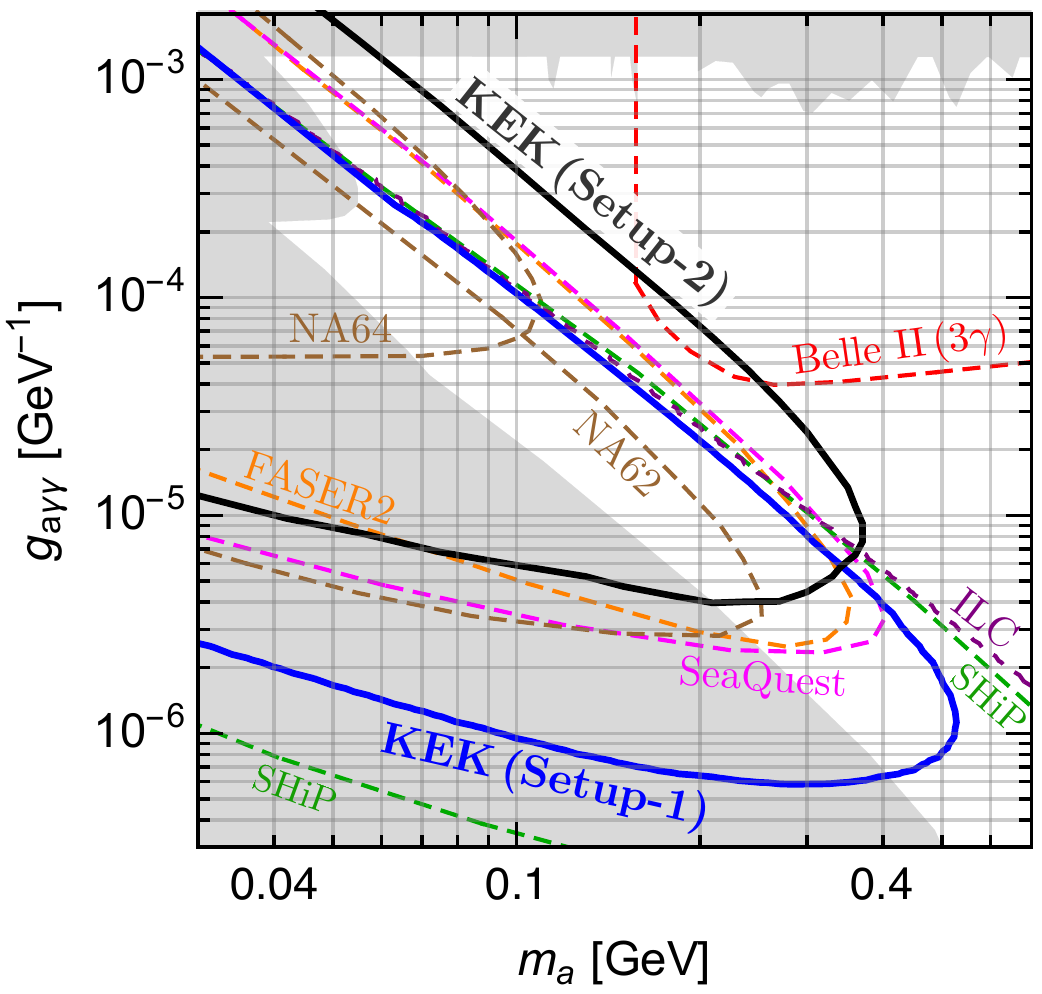}
~~~~~
\includegraphics[width=7.0cm, bb=0 0 300 286]{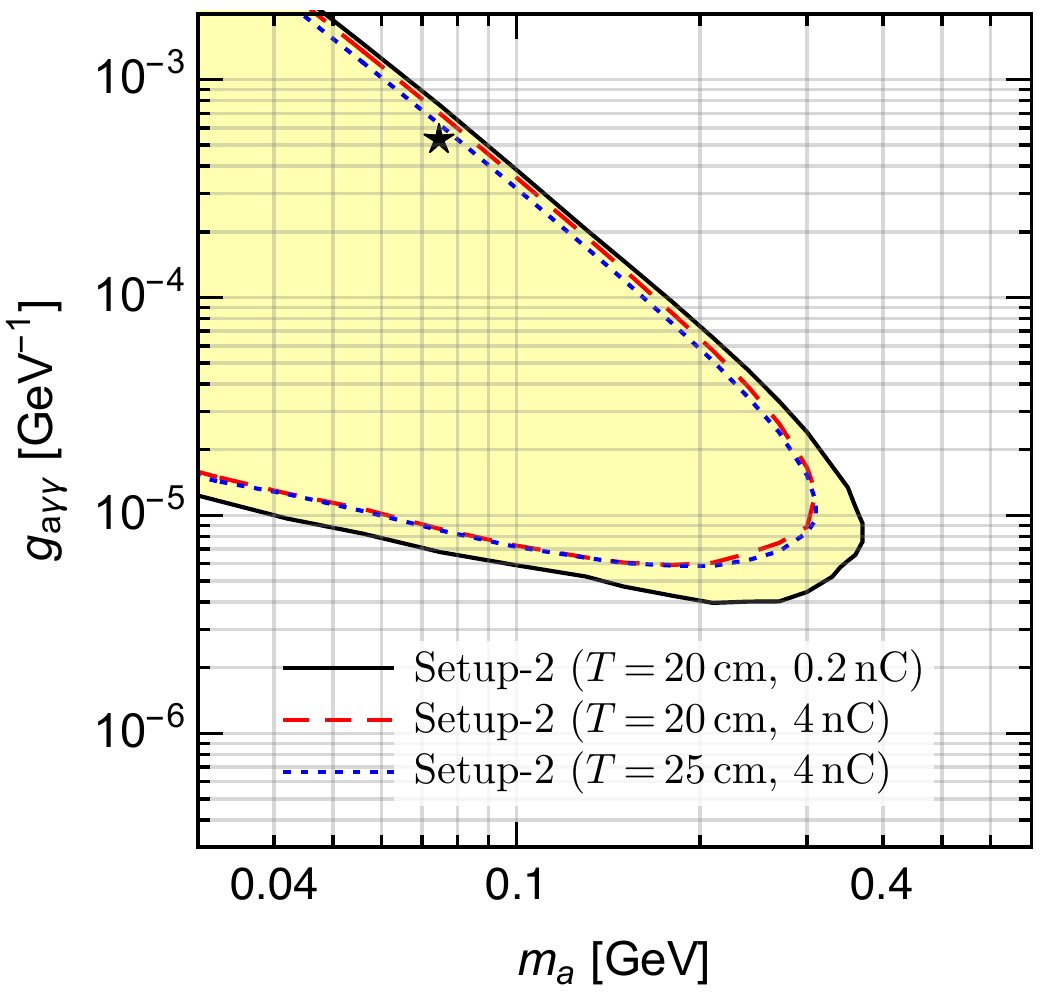}
\caption{{\footnotesize 
{\bf [Left panel]}: 
The blue and black lines show the 95\% C.L. sensitivity to ALP with Setup-1 and Setup-2, respectively. 
For each setup, the number of injected particles is assumed to be $6.5\times10^{18}$, which is equal to the number of particles using the beam current at KEK Linac for one month. For Setup-2, the bunch charge is assumed to be 0.2\,nC. 
The dashed lines show the sensitivity in other beam dump and forward detector experiments, 
    NA62~\cite{NA62:2017rwk,Dobrich:2015jyk}, 
    NA64~\cite{Dusaev:2020gxi}, 
    SHiP~\cite{Dobrich:2015jyk,Dobrich:2019dxc}, 
    FASER~\cite{Feng:2018pew}, 
    SeaQuest~\cite{Berlin:2018pwi}, 
    ILC~\cite{Sakaki:2020mqb}, 
and from the $e^+e^-$ collider experiment Belle~II~\cite{Dolan:2017osp}.
The gray shaded region shows the region excluded by the previous studies~\cite{
Bjorken:1988as, Abbiendi:2002je, Knapen:2016moh, Dobrich:2015jyk, Aloni:2019ruo, Banerjee:2020fue, BelleII:2020fag}. 
{\bf [Right panel]}: 
The change in sensitivity with different bunch charges and the thickness of the Front shield in Setup-2. 
The solid line is the same as the result of the left panel. 
The red dashed line shows the case where the bunch charge is set to 4\,nC which is the maximum value used for SuperKEK. 
The blue dotted line is for a 4\,nC bunch with a shield thickness of 25\,cm. 
The signal parameter used in Fig.~\ref{fig:mass} is marked with a star symbol.}
}
\label{fig:ALP}
\end{center}
\end{figure}

In Fig.\,\ref{fig:ALP} (left), 
the blue and black lines show the 95\% C.L. sensitivity to ALP in Setup-1 and -2. 
For each setup, the number of injected positrons or electrons is assumed to be $6.5\times10^{18}$, which corresponds to one month of operation with the beam current at KEK Linac. 
For Setup-2, the bunch charge is assumed to be 0.2\,nC. 
We do not mention the bunch charge for Setup-1 since the sensitivity of Setup-1 is independent of the bunch charge. 
The dashed lines show the sensitivity from other beam dump and forward detector experiments, 
    NA62~\cite{NA62:2017rwk,Dobrich:2015jyk}, 
    NA64~\cite{Dusaev:2020gxi}, 
    SHiP~\cite{Dobrich:2015jyk,Dobrich:2019dxc}, 
    FASER~\cite{Feng:2018pew}, 
    SeaQuest~\cite{Berlin:2018pwi}, 
    ILC~\cite{Sakaki:2020mqb}, 
and from the $e^+e^-$ collider experiment Belle II~\cite{Dolan:2017osp}.
The gray shaded region shows the constraint from the previous studies~\cite{
Bjorken:1988as, 
Abbiendi:2002je, 
Knapen:2016moh,  
Dobrich:2015jyk, 
Aloni:2019ruo,   
Banerjee:2020fue, 
BelleII:2020fag 
}.

Setup-1 has better sensitivity to larger-mass and smaller-coupling (lower right direction on the plot) than the previous experiments and some future experiments.  
The number of incident particles is important for improving sensitivity in this direction, therefore SHiP and ILC beam dump experiments, which use a larger number of incident particles, are more sensitive to this region than Setup-1. 
We, however, emphasize the advantage of Setup-1 which can be achieved much earlier and with a much lower cost than SHiP and ILC beam dump experiments.
Setup-2 has a better sensitivity to larger-mass and larger-coupling regions compared to other beam dump and forward detector experiments.

Figure \ref{fig:ALP} (right) shows the variation of sensitivity with different bunch charge and thickness of Front shield in Setup-2.
The solid line is the same as the result of Setup-2 in Fig.\,\ref{fig:ALP} (left). 
The red dashed line shows the case where the bunch charge is set to 4\,nC which is the maximum value used for SuperKEK.
The sensitivity decreases with higher bunch charge due to increase of the pileup.\footnote{With $T=20$\,cm and 4\,nC bunch charge, it might be difficult to search ALP with small mass $(m_a<0.04\,{\rm GeV})$ and large coupling $(g_{a\gamma\gamma} \gtrsim 10^{-3}/{\rm GeV})$. 
This is because the low-energy photons leaking from electromagnetic showers per bunch becomes sizable in the signal region of this parameter space, which is the very forward region $(<10\,{\rm mrad})$. 
The energy deposit in the detector by the leakage photons depends on its configuration and detailed study is left as a subject for future work.}
The blue dotted line is for a 4\,nC bunch with a shield thickness of 25\,cm. 
It can be seen that the upper boundary of the contour shrinks, compared to the case of $T=20\,$cm.  This is because the decay length of ALP in that parameter region is only a few cm, and a longer shield increases the probability of ALP decay in the shield and reduces the number of signals.
We note that the signal parameter in Fig.~\ref{fig:mass} is marked with a star symbol.

\section{Summary}
We proposed a fixed-target experiment using electron and positron beams available at KEK Linac with two shielding setups (Setup-1 and -2) to search for ALPs coupling to photons. These setups are designed to increase sensitivity to ALP with a shorter lifetime than the previous beam dump experiments. Setup-1 is designed to uses a 2\,m shield, while Setup-2 uses a 20-25\,cm shield with an additional hollow shield.

The sensitivity of these setups to ALP was evaluated and compared with that in the other beam dump experiments, fixed-target experiments, and $e^+e^-$ collider experiments. 
We found that the setups are highly sensitive to the gap between the regions explored by the previous beam dump experiments and the Belle II experiment.

\subsection*{Acknowledgement}
We would like to thank Hiroshi Iwase, Takahiro Oyama, Hirohito Yamazaki, Mihoko M. Nojiri, Mitsuhiro Yoshida, Shinichiro Abe, Hirokazu Ikeda and Toru Takeshita for helpful discussions and comments. This work was supported by JSPS KAKENHI Grant Number JP20K04004 and JP21H00082.

\bibliographystyle{unsrt}
\bibliography{refs}

\end{document}